# Enhancing Reasoning Capacity of SLM using Cognitive Enhancement

Jonathan Pan, Swee Liang Wong, Xin Wei Chia, Yidi Yuan
Home Team Science and Technology Agency, Singapore
Jonathan_Pan@htx.gov.sg, Wong_Swee_Liang@htx.gov.sg, Chia_Xin_Wei@htx.gov.sg, Yuan_Yidi@htx.gov.sg

*Abstract*— Large Language Models (LLMs) have been applied to automate cyber security activities and processes including cyber investigation and digital forensics. However, the use of such models for cyber investigation and digital forensics should address accountability and security considerations. Accountability ensures models have the means to provide explainable reasonings and outcomes. This information can be extracted through explicit prompt requests. For security considerations, it is crucial to address privacy and confidentiality of the involved data during data processing as well. One approach to deal with this consideration is to have the data processed locally using a local instance of the model. Due to limitations of locally available resources, namely memory and GPU capacities, a Smaller Large Language Model (SLM) will typically be used. These SLMs have significantly fewer parameters compared to the LLMs. However, such size reductions have notable performance reduction, especially when tasked to provide reasoning explanations. In this paper, we aim to mitigate performance reduction through the integration of cognitive strategies that humans use for problem-solving. We term this as cognitive enhancement through prompts. Our experiments showed significant improvement gains of the SLMs' performances when such enhancements were applied. We believe that our exploration study paves the way for further investigation into the use of cognitive enhancement to optimize SLM for cyber security applications.

## I. Introduction

In the field of cyber investigation and digital forensics, the analysis of logs is a frequent activity. It is also an important research topic with practical significance in the field of failure identification [1], [2] and security threat detection [3], [4]. Such analysis is done to facilitate the detection of anomalous activities so that immediate or corresponding remediation may be done to contain or remediate the issue recorded in the logs. The issue may affect system resiliency against system faults, degradation and intentionally induced cyber physical attacks. However, the analysis of logs has its complexities namely from being voluminous, varied, and contextual. Additionally, logs possess inherent semantic complexity [5].

With the recent advances with Large Language Models (LLMs), they provide the means to automate the analysis of logs [6][7][8][9]. While automation provides productivity gains, explain-ability of the analysis and conclusions made by such Artificial Intelligence (AI) solutions remains an important attribute for such solutions to be adopted [10]. Another issue with LLMs is that not all digital forensics or investigative activities are suitable for services delivered through online channels[11]. A locally deployed model would be needed where there are privacy and confidentiality concerns. However, such online LLMs are measurably larger with their resource demands requiring more compute resources in the form of GPUs and memory. An alternative solution is to run smaller versions of their large contemporaries on local compute resources like GPU equipped laptops, work stations or servers. In this research work, we explore the use of cognitive enhancement through prompts to improve the performance of the Smaller Large Language Models (SLMs).

In the next section, we will cover the complexity of performing log analysis and considerations when using language models for such task. This is followed by a review of current research work in the use of language models for log analysis and ways to improve the performances of SLMs. We then described our proposed cognitive enhancement technique used with details of the experimental setup and its evaluation. This paper concludes with a summary of this work and potential future research direction.

## II. Background Information

In this section, we articulate the background information related to the complexity in performing log analysis. We also consider what is needed if language models are practically used for such task.

### A. Complexity with Log Analysis

The form for logs is typically unique to how the software has been developed or configured to post entries into these textual files. These logs are contextual to the environment which the system resides in [5]. Hence, the analysis of such log datasets requires contextual understanding of the system or component that generates such logs especially when attempting to classify or distinguish what is a normal log entry and what is not. Anomaly recordings in logs may not include explicit keywords like 'Error' or 'Failure' to which signature-based rule engine could detect and draw needed attention. The limited sample size of the varied forms of anomalous log entries would create more constraints to enable the development of a robust machine learning model to detect anomalies from logs. Hence log analysis requires semantic comprehension [12][5] and to overcome the constraints of having limitedly available information about the form of anomalies that could occur.

*B. Consideration on the Use of Language Models*

In our previous research work, we demonstrated that we could use a LLM to perform log anomaly detection with a vector database containing only selected samples of normal log entries based on the log entries clustering distribution. Our model construct, that we called RAGLog [9] used the Retrieval Augmented Generative approach with GPT 3.5 to perform logs analysis that performed relatively well for a zero-shot classifier.

However, for cybersecurity applications, we will require the construct to work on locally deployed LLM instead of using online LLM instances. This is to address concerns with privacy and confidentiality. However, locally deployed LLM have limited resources available in terms of compute and memory resources. Hence, local deployment would be limited to SLMs to operate within such constraints. Given the absence of clear technical definitions of SLMs in the field, in this paper, we describe them as models that can be readily operated on a consumer hardware. This includes models such as LLaMa 2 7b. However, it is known that these smaller models have comparatively limited capabilities as compared to their larger counterparts. In this research context, the smaller models have difficulties solving complex tasks that involves multiple reasoning steps [13]. This research work seeks to deal with that.

Another important consideration with the use of LLM for log analysis and for the broader cyber security application is to have the means for humans to evaluate the correctness of the models' assessment. This will help the human log reviewer or investigator to assess whether the anomaly classification is valid and warrants further investigation. In the absence of the means to evaluate the correctness of such models, these cyber security practitioners would be less inclined to accept the decisions made by these models[14].

III. RELATED WORK

In this section, we review the current state of art research work in the use of Language Models to perform log analysis and log anomaly detection. We also review how Small Language Models are being tuned for high performance.

*A. Language Models for Log Analysis*

There are several research attempts to apply Large Language models to perform log analysis. Qi et al. [6] proposed a framework for log-based anomaly detection using ChatGPT using varied prompt constructs, window sizes and input sequences. Their work showed the non-triviality of an optimal prompt, window size limitations as well as high false positive rates. Mudgal et al. [7] designed specific prompts with ChatGPT for log parsing that had excellent performance. However, with other areas of log analysis like anomaly detection and log summarization, the LLM exhibited limitations that warrant further research. Liu et al. [8] tested their LogPrompt model in zero-shot scenarios with varying number of provided log samples and different prompt formats (self-prompt, CoT prompts and In-context Prompt). The zero-shot test results showed promise when compared with our log analysis algorithms and other Deep Learning architectures. However, it had very low precision scores which is not optimal if applied to log analysis for operations and maintenance activities to support resiliency. We also applied GPT 3.5 to perform log anomaly detection using Retrieval Augmented Generation approach which we called RAGLog [9]. However, all such approaches focused on the use of Large Language Models instead of Small Language Model. This research work seeks to address this gap.

*B. Improving Smaller Large Language Models*

There are a number of studies in how to teach or improve the capabilities of small language models. Magister et. al [15] argued that one approach, through the use of knowledge distillation, enables the transfer of reasoning capabilities from a large model of over 100 billion parameters to smaller models. Such finetuning approach improves the task accuracy of these smaller models across a range of benchmarking datasets. Xie et. al [29] suggested training several small language models with multiple candidates plan and subsequently select a good one. Another work suggested parameter editing methods and saw performance in small language models [30]. Majority of approaches look to improve SLMs through fine tuning [31]. However, this approach requires expertise and computational resources, which poses a challenge for local deployment where resources are limited. Therefore, there is a need to explore alternative strategies to improve the performance of SLMs without the need of extensive computational resources.

*C. Cognitive Enhancements of Language models*

In this paper, we explore the use of cognitive enhancements to improve the outcome of a SLM. While there has been extensive research into the use of prompt engineering to improve performance, such as Chain-of-Thought (CoT) [34] and Self-Consistency [33] these approaches were often performed on larger language models. Importantly, it was found that the use of CoT only yielded performance improvement when used on models with larger than 100B model parameters. Smaller models produced illogical outcomes, leading to poorer outcomes. The same limitations were observed for Self-Consistency where the gain in accuracy was lower for smaller models. Inspired by introspective reasoning in human cognition, Wang and Zhao [35] explored the use of metacognition to improve the outcome of LLMs. Metacognitive prompting led to an improvement of accuracy when compared to standard and CoT prompting. However, like CoT, performance gains were only observed in the larger GPT-4 model. Hence, there is a need to further investigate cognitive enhancement techniques that would significantly improve the performance of SLMs.

IV. COGNITIVE ENHANCEMENT

Our research question attempts to address the question whether smaller language models, with its comparatively weaker cognitive capabilities, could improve its performance through cognitive enhancements. More specifically, we hypothesized that decomposing a task into smaller manageable steps can improve the outcome of SLMs on a combined reasoning and decision-making or classification task. We also applied Self-Reflection to validate the reasoning and decision-making cognitive process.

*A. Task Decomposition*

In the study of psychology and cognitive science, there are studies on the effects of increased levels of cognitive load that could cause people to make poorer decisions [19]. One approach proposed to deal with cognitive overload is the use of task decomposition [20]. Task decomposition is the process of breaking down complex tasks or problems into smaller manageable tasks or steps. It also involves defining the sequence of subgoals to achieve the main goal or objective [21][22]. We propose that we could apply the same approach to break down the complex task of performing log analysis into smaller sub-tasks that the small language models are better capable of handling and to assess whether our approach leads to better and comparable performances against their much larger counterparts. There exists past work in the use of Task Decomposition [23]. However, they are focused on its application in LLMs to enhance their performance instead of SLMs.

We propose the following mathematical formulation for task decomposition in Equation 1, where a complex composite task $c$ is broken into simpler sequentially concatenated tasks $c_i$. For each decomposed task an input $x_i$ formulated in the form of a prompt is given to the language model. The reply from the language model would be the output $y$ to the prompt query and classification conclusion to the query task. When there is a concatenation of two cognitive tasks with the additive operation, the output of the preceding cognitive task would be the input to the proceeding cognitive task (Equation 2). This continues until the entire set of the decomposed tasks is done. In contrast, without task decomposition, a composite task can be given an input in the form of a composite input $x$ (Equation 3).

$$c = c_0 + c_1 \quad (1)$$

$$y = c_0(x_0) + c_1(x_1) \quad (2)$$

$$y = c(x) \quad (3)$$

An example of the prompt-based task decomposition is as such.

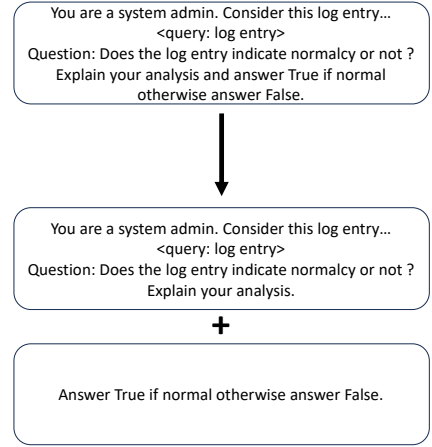

Figure 1. Task Decomposition illustrating the decomposition of a prompt into two sequential Explain and Decide prompts respectively.

*B. Self-Reflection*

As the analysis task was decomposed into smaller tasks namely with the reasoning (Explaining) and classification (Deciding) as two smaller tasks, we included self-reflective task. The self-reflective task acts as a feedback activity for the language model to validate the earlier reasoning and classification conclusion. Shinn et. al [32] argues that using self-reflection improves decision-making performance.

V. METHODOLOGY AND ANALYSIS

In our experiment setup, we designed our experiment to address our research question whether Retrieval Augmented Generation in LLM could perform log anomaly detection.

*A. Log Datasets*

For our log datasets, we used BGL [25] and Thunderbird [26]. These are two popular datasets typically used by researchers to evaluate log analysis models [24].

The BGL are open real-world datasets from HPC from a BlueGene/L supercomputer at Lawrence Livermore National Labs. This dataset has an important characteristic associated with their appearance of many new log messages in the timeline of the data, that is, the systems change over time. The Thunderbird open dataset of logs was collected by Sandia National Lab. It contains alert and non-alert messages. Both datasets are labelled with sizeable imbalance for the anomaly class.

*B. Evaluation Metrics*

As the dataset used had binary classification labels, we used *Precision* to measure the accuracy of the model against type I error (true positive) and *Recall* to measure the accuracy of the models against type II error (true negative). Finally, we used *F1 score* to measure the harmonic mean of *precision* and *recall*.

$$Precision = \frac{TP}{TP + FP} \quad (4)$$

$$Recall = \frac{TP}{TP + FN} \quad (5)$$

$$F1\ score = 2 \times \frac{Precision \times Recall}{Precision + Recall} \quad (6)$$

TP (True Positive) represents the number of correctly classified anomalies, TN (True Negative) represents normal log entries and FP (False Positive) is the number of incorrect anomaly classification. FN (False Negative) is the number of incorrect classifications of log entries as normal while the label or ground truth states overwise.

*C. Experimentation Preparation and Execution*

We first populated the vector database with selected samples and limited size of the log database with normal log entries. The selection was done by first applying unsupervised k-means clustering to the dataset and populating the database from random sampling from the cluster classes. This is to ensure that the sample selection had a good distribution of the normal log entries.

For our experiment, we used four variations of pretrained Small Language Models. They are Meta's LLaMa 2 7B, Meta's LLaMa 2 13B, Vicuna 7B and Vicuna 13B. The LLaMa 2 [27] is an auto-regression language model that uses an optimized transformer architecture. The models are tuned using supervised fine-tuning (SFT) and reinforcement learning with human feedback (RLHF) to align to human preferences for helpfulness and safety. The Vicuna models [28] are fine-tuned from LLaMa 2 using high quality conversations using user-shared conversations hosted in ShareGPT.com. Due to resource limitations, we used the quantized versions of the models specifically the GGUF Q4_K_M format by llama.cpp team.

The experiment configuration for this research work used the the same construct as our previous experiment in our work in RAGLog [9]. We used vector databases to store the embeddings of limited sized selected log entries. The log entries in the database contain only the normal log entries and number no more than two thousand. The selection of the normal log entries to be populated is done first with the clustering of normal log entries and random selection of cluster classes. Using the Retrieval Augmented Generation approach [17], the Small Language Model would be queried with the log entry under evaluation using the Question and Answer template [18]. The orchestrator would then query the vector database through the embedding model to retrieve the best matched normal log entry. The resultant from the vector database would then be given to the SLM with our specially constructed prompts with the final concluding reply provided by the orchestrator where ground truth verification is done against the provided labels for corresponding dataset.

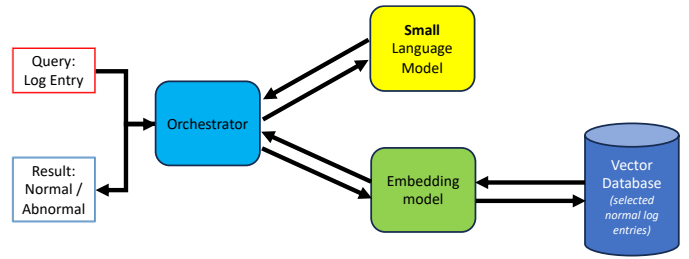

Figure 2. Workflow of log analysis.

The concluding reply to the queried log entry would contain the reasoning explanation to facilitate validation by a human analyst and the inferred classification conclusion that would be used to assess the model's performance.

For this paper, the orchestrator was scripted to enable the incorporation of cognitive enhancement. The main variable to our research construct is the inclusion of Task Decomposition through prompts. For our experiment, we used two such forms namely Explain First and Decide Later or Decide First and Explain Later. The Explain First and Decide Later is an approach proposed by Yao et. al [16] with their ReAct framework for LLMs to generate reasoning traces and task-specific actions. Results from research work showed that their framework outperform several state-of-the-art baseline tests on decision-making tasks. As part of this experiment, we attempted to explore the alternative, Decide First and Explain Later to investigate if the sequence of task decomposition affects performance. We included Self-Reflection into our cognitive process as the concluding cognitive process task after both reasoning explanation and classification conclusion were generated.

Hence, our experiments conducted were as such with the corresponding prompts.

- [{E,D}+R] Explain and Decide then Reflect; where Explain and Decide are done as one activity without task decomposition followed by self-reflection.

- [{D,E}+R] Decide and Explain then Reflect; where Decide and Explain are done as one activity without task decomposition followed by self-reflection

- [E+D+R] Explain, Decide then Reflect; where each activity is done in the mentioned sequence with task decomposition followed by self-reflection.

- [D+E+R] Decide, Explain then Reflect; where each activity is done in the mentioned sequence with task decomposition followed by self-reflection.

Each activity was executed with its corresponding prompt. One prompt was used for the combined pair of Decide and Explain or Explain and Decide when they were executed as one activity. The reply responses from the models following the query prompts were evaluated against their corresponding labels (normal or anomaly).

## D. Results and Analysis

The following are our experiment test results for both datasets (BGL and Thunderbird). Bolded scores indicate the best result for each model.

| Model | w/o Task Decomposition | | w/ Task Decomposition | |
|---|---|---|---|---|
| | {E,D}+R | {D,E}+R | E+D+R | D+E+R |
| LLaMA 2 7B | 0.74 | 0.67 | 0.71 | **0.82** |
| LLaMA 2 13B | 0.71 | 0.67 | **0.72** | 0.58 |
| Vicuna 7B | 0.67 | 0.71 | 0.72 | **0.75** |
| Vicuna 13B | 0.69 | 0.67 | 0.80 | **0.86** |

Table 1. F1 scores for experiments involving BGL dataset

| Model | w/o Task Decomposition | | w/ Task Decomposition | |
|---|---|---|---|---|
| | {E,D}+R | {D,E}+R | E+D+R | D+E+R |
| LLaMA 2 7B | 0.68 | 0.70 | **0.75** | 0.74 |
| LLaMA 2 13B | 0.82 | 0.72 | 0.88 | **0.94** |
| Vicuna 7B | 0.68 | 0.70 | **0.76** | 0.73 |
| Vicuna 13B | 0.71 | 0.68 | 0.90 | **0.94** |

Table 2. F1 scores for experiments involving Thunderbird dataset

The highlighted F1 scores showed significant improvements especially when Task Decomposition was applied. However, we do notice that for conditions LLaMa 2 7B with [E+D+R] and LLaMA 2 13B with [D+E+R], Task Decomposition led to a reduction in performance when compared to their respective conditions without Task Decomposition. This warrants further experiments especially with the tuning of prompts. In addition, we found that the sequence of Explain and Decide does not play a significant role in the model's performance.

## VI. CONCLUSION AND FUTURE DIRECTIONS

Our research work explored the use of Smaller Large Language Model (SLM) to perform log anomaly detection based on our earlier work that applied the Retrieval Augmented Generation approach with GPT 3.5 (RAGLog). As SLMs lack reasoning capacity in comparison to LLMs, we used Task Decomposition to decompose the prompt into smaller manageable steps and noted significant performance improvements. By applying this to a log analysis cybersecurity application, we improved the performance of SLMs to achieve robust outcomes while simultaneously addressing concerns related to data privacy and confidentiality. Our study demonstrates the plausible use of such cognitive enhancement through Task Decomposition to improve SLMs' capabilities.

The next step to this research work is to apply this to real-world log analysis and assess the performance. Also, we will look into how such language models can reduce the time needed for such log analysis as the current approach performs log analysis sequentially.

## REFERENCES

[1] A. Pecchia, D. Cotroneo, Z. Kalbarczyk, and R.K. Iyer, "Improving log-based field failure data analysis of multi-node computing systems", DSN'11: Proc. of the 41st IEEE/IFIP International Conference on Dependable Systems and Networks, pages 97–108. IEEE, 2011.

[2] W. Xu, L. Huang, A. Fox, D. Patterson, and M.I. Jordon, "Detecting large-scale system problems by mining console logs", SOSP'09: Proc. of the ACM Symposium on Operating Systems Principles, 2009.

[3] A. Brandao and P. Georgieva, "Log Files Analysis For Network Intrusion Detection," 2020 IEEE 10th International Conference on Intelligent Systems (IS), 2020, pp. 328-333, doi: 10.1109/IS48319.2020.9199976.

[4] M. Moh, S. Pininti, S. Doddapaneni and T. Moh, "Detecting Web Attacks Using Multi-stage Log Analysis," 2016 IEEE 6th International Conference on Advanced Computing (IACC), 2016, pp. 733-738, doi: 10.1109/IACC.2016.141.

[5] A. Ekelhart, E. Kiesling and K. Kurniawan, "Taming the logs – Vocabularies for semantic security analysis", SEMANTiCS 2028 – 14th International Conference on Semantic Systems, Science Direct, Procedia Comput Science 137, pp. 109-119, 2018.

[6] J. Qi, S. Huang, Z. Luan, C. Fung, H. Yang and D. Qian, "LogGPT: Exploring ChatGPT for Log-Based Anomaly Detection", arXiv:2309.01189v1, 2023.

[7] P. Mudgal and R. Wouhaybi, "An Assessment of ChatGPT on Log Data", arXiv:2309.07938v1, 2023.

[8] Y. Liu, S. Tao, W. Meng, J. Wang, W. Ma, Y. Zhao, Y. Chen, H. Yang, Y. Jiang and X. Chen, "LogPrompt: Prompt Engineering Towards Zero-Shot and Interpretable Log Analysis", arXiv:2308.07610v1, 2023.

[9] J. Pan, S.L. Wong and Y. Yuan, "RAGLog: Log Anomaly Detection using Retrieval Augmented Generation", arXiv, arXiv:2311.05261, https://doi.org/10.48550/arXiv.2311.05261.

[10] Z. Zhang, H. A. Hamadi, E. Damiani, Y. Y. Chan and F. Tahar, "Explainable Artificial Intelligence Applications in Cyber Security: State-of-the-Art in Research", IEEE Access, pp. 93104-93139, 2022, https://doi.org/10.1109/ACCESS.2022.3204051.

[11] M. Scanlon, F. Breitinger, C. Hargreaves, J. Hilgert and J. Sheppard, "ChatGPT for digital forensic investigation: The good, the bad and the unknown", Forensic Science International: Digital Investigation, Vol. 46 Supplement, Oct 2023, https://doi.org/10.1016/j.fsidi.2023.301609.

[12] V. H. Le and H. Zhang, "Log-based Anomaly Detection without Log Parsing", 2021 36th IEEE/ACM International Conference on Automated Software Engineering (ASE), Nov 2021.

[13] W. X. Zhao, K. Zhou, J. Li, T. Tang, X. Wang, Y. Hou, Y. Min, B. Zhang, J. Zhang, Z. Dong, Y. Du, C. Yang, Y. Chen, Z. Chen, J. Jiang, R. Ren, Y. Li, X. Tang, Z. Liu, P. Liu, J. Y. Nie and J. R. Wen, "A Survey of Large Language Models", arXiv, arXiv:2303.18223, https://doi.org/10.48550/arXiv.2303.18223,

[14] N. Capuano, G. Fenza, V. Loia and C. Stanzione, "Explainable Artificial Intelligence in CyberSecurity: A Survey", IEEE Access, vol. 10, pp. 93575-93600, Sep 2022.

[15] L. C. Magister, J. Mallinson, J. Adamek, E. Malmi and A. Severyn, "Teaching Small Language Models to Reason", arXiv, arXiv:2212.08410, https://doi.org/10.48550/arXiv.2212.08410.

[16] S. Yao, J. Zhao, D. Yu, N. Du, I. Shafran, K. Narasimhan and Y. Cao, "ReAct: Synergizing Reasoning and Acting in Language Models", arXiv, arXiv:2210.03629, https://doi.org/10.48550/arXiv.2210.3629.

[17] P. Lewis, E. Perez, A. Piktus, F. Petroni, V. Karpukhin, N. Goyal, H. Küttler, M. Lewis, W. Yih, T. Rocktäschel, et al., "Retrieval-augmented generation for knowledge-intensive nlp tasks", Advances in Neural Information Processing Systems, 33:9459–9474, 2020.

[18] K. Lee, M. W. Chang, and K. Toutanov, "Latent retrieval for weakly supervised open domain question answering", arXiv:1906.00300, 2019.

[19] C. Deck and S. Jahedi, "The effect of cognitive load on economic decision making: A survey and new experiment", European Economic Review, vol. 78, pp. 97-119, Aug 2015, https://doi.org/10.1016/j.euroecorev.2015.05.004.

[20] B. M. Knisely, J. S. Joyner, A. M. Rutkowski, M. Wong, S. Barksdale, H. Hotham, K. Kharod and M. Vaughn-Cooke, "A cognitive decomposition to empirically study human performance in control room environments", International Journal of Human-Computer Studies, vol. 141, Sep 2020, https://doi.org/10.106/j.ijhcs.2020/102438.

[21] C. G. Correa, M. K. Ho, F. Callaway and T. L. Griffiths, "Resource-rational Task Decomposition to Minimize Planning Costs", 42nd Annual Meeting of the Cognitive Science Society: Developing a Mind: Learning in Humans, Animals, and Machines, CogSci 2020, pp. 2974-2980, 2020.

[22] J. M. Schraagen, S. F. Chipman and V. L. Shalin, "Cognitive Task Analysis", Psychology Press, 2000.

[23] X. Huang, W. Liu, X. Chen, X. Wang, H. Wang, D. Lian, Y. Wang, R. Tang and E. Chen, "Understanding the planning of LLM agents: A survey", arXiv, arXiv:2402.02716, https://doi.org/10.48550/arXiv.2402.02716.

[24] Z. Chen, J. Liu, W. Gu, Y. Su, and M. R. Lyu, "Experience Report: Deep Learning-based System Log Analysis for Anomaly Detection,", arXiv, arXiv:2107.05908, https://doi.org/10.48550/arXiv.2107.05908.

[25] A. Oliner and J. Stearley, "What supercomputers say: A study of five system logs", 37th Annual IEEE/IFIP International Conference on Dependable Systems and Networks (DSN'07). IEEE, pp 575–584, 2007.

[26] A. Oliner and J. Stearley, "What supercomputers say: A study of five system logs," in DSN, 2007.

[27] H. Touvron et. al, "LLAMA 2: Open Foundation and Fine-Tuned Chat Models", arXiv, arXiv:2307.09288, https://doi.org/10.48550/arXiv.2307.09288.

[28] L. Zheng, W.L. Chiang, Y. Sheng, S. Zhuang, Z. Wu, Y. Zhuang, Z. Lin, Z. Li, D. Li, E. P. Xing, H. Zhang, J. E. Gonzalez and I. Stoica, "Judging LLM-as-a-Judge with MT-Bench and Chatbot Arena", arXiv, arXiv:2306.05685, https://doi.org/10.48550/arXiv.2306.05685.

[29] S. M. Xie, H. Pham, X. Dong, N. Du, H. Liu, Y. Lu, P. Liang, Q. V. Le, T. Ma and A. W. Yu, "Doremi: Optimizing data mixtures speeds up language model pretraining", arXiv, arXiv:2305.10429.2023.

[30] Y. Yao, P. Wang, B. Tian, S. Cheng, Z. Li, S. Deng, H. Chen and N. Zhang, "Editing large lanague models: Problems, method and opportunities", CoRR, vol. abs/2305.13172, 2023.

[31] J. Devlin, M. Chang, K. Lee, and K. Toutanova, "BERT: pre-training of deep bidirectional transformers for language understanding," in Proceedings of the 2019 Conference of the North American Chapter of the Association for Computational Linguistics: Human Language Technologies, NAACL-HLT 2019, Minneapolis, MN, USA, June 2-7, 2019, Volume 1 (Long and Short Papers), J. Burstein, C. Doran, and T. Solorio, Eds. Association for Computational Linguistics, 2019, pp. 4171–4186.

[32] N. Shinn, B. Labash, and A. Gopinath. Reflexion: an autonomous agent with dynamic memory and self-reflection. arXiv preprint arXiv:2303.11366, 2023.

[33] X. Wang, J. Wei, D. Schuurmans, Q. Le, E. Chi, S. Narang, A. Chowdhery and D. Zhou, "Self-Consistency Improves Chain of Thought Reasoning in Language Models", arXiv preprint arXiv:2203.11171, 2023.

[34] J. Wei, X, Wang, D. Schuurmans, M. Bosma, B. Ichter, F. Xia, E. Chi, Q. Le and D. Zhou, "Chain-of-Thought Prompting Elicits Reasoning in Large Language Models", arXiv preprint arXiv:2201.11903, 2023.

[35] Y. Wang and Y. Zhao, "Metacognitive Prompting Improves Understanding in Large Language Models", arXiv preprint arXiv:2308.05342v4, 2024.